# Estimating Total Lung Volume from Pixel-level Thickness Maps of Chest Radiographs Using Deep Learning

Original Research - Published in the Journal of Radiology: Artificial Intelligence

(DOI: 10.1148/ryai.240484)

Tina Dorosti*, Manuel Schultheiß*, Philipp Schmette, Jule Heuchert, Johannes Thalhammer, Florian T. Gassert, Thorsten Sellerer, Rafael Schick, Kirsten Taphorn, Korbinian Mechlem, Lorenz Birnbacher, Florian Schaff, Franz Pfeiffer, Daniela Pfeiffer

## Key points:

1. Total lung volume (TLV) predictions by the U-Net deep learning model showed low error rates for synthetic (MSE = 0.20 $L^2$) and real (MSE = 0.35 $L^2$) radiographs from the Klinikum Rechts der Isar dataset.
2. TLV estimated from synthetic and real chest radiographs by the model using lung thickness maps strongly correlated with CT-derived TLV ($r = 0.97$ and 0.91, respectively; both $p < .001$).
3. The U-Net model demonstrated robust performance across different datasets, with the lowest mean squared error (MSE = 0.09 $L^2$) and highest correlation ($r = 0.99$, $p < .001$) achieved on the Luna16 test data for TLV estimation.

## Summary statement:

**A deep learning model successfully estimated lung thickness maps and total lung volumes from chest radiographs, demonstrating strong correlation with CT-derived measurements.**

**Abbreviations:** Chronic Obstructive Pulmonary Disease (COPD), Chest Radiograph (CXR), Hounsfield Units (HU), Mean Squared Error (MSE), Mean Absolute Error (MAE), Mean Absolute Percentage Error (MAPE), Pulmonary Embolism (PE), Total Lung Volume (TLV)

## Abstract

**Purpose:** To estimate the total lung volume (TLV) from real and synthetic frontal chest radiographs (CXR) on a pixel level using lung thickness maps generated by a U-Net deep learning model.

**Methods:** This retrospective study included 5,959 chest CT scans from two public datasets: the lung nodule analysis 2016 (n=656) and the Radiological Society of North America (RSNA) pulmonary embolism detection challenge 2020 (n=5,303). Additionally, 72 participants were selected from the Klinikum Rechts der Isar dataset (October 2018 to December 2019), each with a corresponding chest radiograph taken within seven days. Synthetic radiographs and lung thickness maps were generated using forward projection of CT scans and their lung segmentations. A U-Net model was trained on synthetic radiographs to predict lung thickness maps and estimate TLV. Model performance was assessed using mean squared error (MSE), Pearson correlation coefficient (r), and two-sided Student's t-distribution.

**Results:** The study included 72 participants (45 male, 27 female, 33 healthy: mean age 62 years [range 34-80]; 39 with chronic obstructive pulmonary disease: mean age 69 years [range 47-91]). TLV predictions showed low error rates ($MSE_{Public\text{-}Synthetic}$=0.16 $L^2$, $MSE_{KRI\text{-}Synthetic}$=0.20 $L^2$, $MSE_{KRI\text{-}Real}$=0.35 $L^2$) and strong correlations with CT-derived reference standard TLV ($n_{Public\text{-}Synthetic}$=1,191, r=0.99, $P$<0.001; $n_{KRI\text{-}Synthetic}$=72, r=0.97, $P$<0.001; $n_{KRI\text{-}Real}$=72, r=0.91, $P$<0.001). The Luna16 test data demonstrated the highest performance, with the lowest mean squared error (MSE = 0.09 $L^2$) and strongest correlation (r = 0.99, $P$ <0.001) for TLV estimation.

**Conclusion:** The U-Net-generated pixel-level lung thickness maps successfully estimated TLV for both synthetic and real radiographs.

## Introduction

Numerous lung conditions, including infectious diseases, interstitial lung diseases, and chronic obstructive pulmonary disease (COPD), impact lung function by altering air volume within the lungs [1–4]. Total lung volume (TLV) is a critical metric for evaluating the severity, progression, and treatment response in cases of restrictive lung diseases. While pulmonary function tests like spirometry and body plethysmography are standard assessment methods, imaging techniques offer an alternative approach, particularly for patients unable to perform pulmonary function tests or when lung imaging is already available [5].

Traditionally, image-based lung volume estimation methods have relied on manual calculations using lateral and frontal chest X-ray radiographs (CXR). Hurtado et al. manually calculated the overall lung area and multiplied it by the PA diameter [2]. Pierce et al. used shape information to gain a more accurate estimate of the volume [6]. Alternatively, X-ray computed tomography (CT) allows for an accurate estimation of the lung volume by multiplying the number of lung area voxels by the voxel size, which itself is determined by the slice thickness and pixel spacing information [5].

Recent advancements in deep learning have demonstrated promising results in lung volume estimation [7-13]: Researchers have developed models combining pulmonary function test results, CT-derived data, as well as CXRs to estimate TLV [7-8, 10, 11]. Boulogne et al. [12] use a deep learning architecture for CT data to predict spirometry and diffusion capacity of carbon monoxide estimates at the patient and lobe levels. Marsh explored the generation of relative lung thickness maps for CT-simulated radiographs [13]. To the best of our knowledge, there are no deep-learning models available for pixel-level lung volume estimation of real CXRs.

Pixel-level thickness maps can provide the location and shape information of healthy and dysfunctional areas across the lung. Such maps can be used to accurately measure volumetric changes in pulmonary tissue, improving risk assessment for pulmonary resection surgery [12, 14]. Furthermore, pixel-level lung volume estimation could be used in novel imaging modalities such as X-ray dark-field imaging, where dividing the measured signal by the lung thickness improves the image contrast [15-16]. More broadly, the proposed method could assist physicians in diagnosing and preventing the progression of pulmonary diseases during routine CXR screenings.

In this exploratory study, we aim to estimate the lung volume on a pixel level using lung thickness maps generated by a U-Net. We simulate synthetic frontal CXRs and two-dimensional lung thickness maps from CT data and volumetric lung segmentations using a forward projection model based on physical processes. This approach ensures applicability across different X-ray machines. We assess the model's performance using both synthetic and real radiographs.

# Methods

The code is available at: https://github.com/tidorosti/lungThickness

## Datasets

For synthetic frontal CXR generation, chest CT scans were retrieved from two publicly available datasets of the lung nodule analysis 2016 (Luna16) challenge [17] (n=888) and the Radiological Society of North America (RSNA) Pulmonary Embolism (PE) detection challenge 2020 [18] (n=7,122). For comparability reasons, only CT scans acquired with 120 kVp were considered, excluding 2,051 scans and resulting in a total of 656 CT scans from the Luna16 dataset and 5,303 CTs from the PE dataset. 94 participants were selected retrospectively (October 2018-December 2019) from the Technical University of Munich's research hospital, Klinikum Rechts der Isar (KRI). Participants scanned without iodinated contrast material (n=3) or with indeterminate diagnosis (n=13) were excluded. Chest CT scans from a total of 72 participants (33 healthy, 39 suffering from COPD) were selected, such that for each participant, a medically indicated frontal CXR acquired in a standing position and no older than seven days was available, excluding an additional 6 participants. Approval from institutional review board was received (IRB code: 87/18 S), and the requirement for written informed consent was waived, as data was analyzed anonymously and retrospectively. The participants in the KRI dataset were previously reported on as part of a prior publication focused on X-ray dark-field chest imaging for assessment of emphysema in patients with COPD [16]. The study included CT data and corresponding CXR attenuation images from these participants to assess the performance of the proposed U-Net on pixel-level lung thickness estimation. CT reconstruction kernel and contrast agent information for all datasets is provided in Table S1 [18][19]. The complete data selection flowchart for both the public and the KRI datasets is shown in Fig. 1.

## Data Preparation

Before the synthetic radiographs were generated from the CT data, the patient table was segmented and removed from the volume. To do so, a binary threshold at -775 Hounsfield unit (HU) was set to separate the air and the body for each CT slice image. An opening filter removed thin lines from partial volume effects between the table and the volume. The torso of the body was selected by finding the biggest connected object with a connected component algorithm. Python's ndimage library from the SciPy package (version 1.10.1) was used [20].

The lung segmentations for the PE and the Luna16 CT data were obtained with the lungmask package (version 0.2.20) [21]. For the KRI data, the automated lung segmentation from CT, as well as reference standard TLV estimations, were supplied by the CT scanner manufacturer (IntelliSpace Portal, Philips Medical Systems, Hamburg, Germany). The slice thickness and pixel spacing of adjacent slices were consistent across each CT scan, ensuring there was neither a gap nor an overlap between the reconstructed slices used for the CT-based reference standard TLV estimations.

## Synthetic Radiograph and Thickness Map Generation

The following method was applied to generate an X-ray spectrum for simulating synthetic radiographs and their corresponding thickness maps via forward projection. An overview of the simulation process is shown in Fig. 2A.

### 2.3.1 X-ray Spectrum Simulation

Certain standard parameters of X-ray systems were set to simulate realistic synthetic radiographs from CT, namely, the tube voltage (70 kVp) and detector properties, such as detector thickness (0.6 mm) and scintillator material (cesium iodide, density=3.38 g/cm$^3$). The tube voltage was selected to match that of the setup used to measure the real radiographs at KRI. The X-ray spectrum was simulated with a semi-empirical model for X-ray transmission using SpekPy (version 2.0.1) [22-24].

### 2.3.2 Material Segmentation

To attribute correct attenuation properties to the different tissue types in the human thorax, the CT scans were segmented into three basis material volumes by applying the following voxel-level thresholds: (-200 – 0) HU for adipose tissue, (0 – 240) HU for soft tissue, and values above 240 HU for bone. These values fell within the ranges described by Buzug et al. [25] and were slightly adjusted to avoid overlapping or missing HU ranges.

To account for differences in material densities, $\rho_i'$, which are reflected in slight variations in the measured HU values, adjusted attenuation values, $\mu_i'(E)$, were calculated for the basis material $i$ in each voxel [26]:

$$\mu_i'(E) = \frac{\mu_i}{\rho_i}(E) \cdot \rho_i'. \tag{1}$$

Here, $\frac{\mu_i}{\rho_i}(E)$ is the tabulated energy-dependent mass attenuation coefficient of basis material $i$, adjusted with the thresholded voxel-wise mass density $\rho_i'$. Considering the definition of the HU value:

$$HU = \frac{\mu - \mu_{water}}{\mu_{water}} \cdot 1000\,, \tag{2}$$

the latter is derived from the measured HU value of material $i$ in a given voxel, $HU_i$, as follows:

$$\rho_i' = \frac{\frac{HU_i}{1000} \cdot \mu_{water}(E_{CT}) + \mu_{water}(E_{CT})}{\frac{\mu_i}{\rho_i}(E_{CT})}\,. \tag{3}$$

The tabulated attenuation coefficients of water, $\mu_{water}(E_{CT})$, and the mass attenuation coefficient of basis material $i$, $\frac{\mu_i}{\rho_i}(E_{CT})$, were retrieved from the NIST database [27] using the xraylib framework (version 4.1.3) [28]. These values were obtained for $E_{CT} = 70 \text{ keV}$, which is the estimated mean energy for the 120 kVp spectrum of the origin CT scanner.

### 2.3.3 Forward Projection

For each CT scan, the density volumes and the corresponding lung segmentation volume were forward-projected to obtain the synthetic radiograph projections and the matching 2-dimensional reference standard lung-thickness map. A cone-beam projector was implemented with the Astra toolbox (version 2.1.3) [29] using the following parameters: detector size = 512x512 pixels, source-to-sample distance = 2,060 mm, and sample-to-detector distance = 375 mm. 10 projections were created for each sample by taking measurements from -10° to 10° at 2° steps.

The intensity value $I$ of each pixel in the synthetic radiograph was calculated with

$$I = \sum_{E=1}^{K} \Phi(E) \cdot \exp\left(\sum_{i=1}^{N} \frac{\mu_i}{\rho_i}(E) \cdot \rho'\right), \tag{4}$$

given the energy-dependent number of photons $\Phi(E)$, a kilo-voltage peak $K$, and $N$ basis materials. Lastly, the negative logarithm of the intensity, normalized with respect to the flat-field image, was taken for a conventional depiction of the radiograph.

## U-Net Implementation

A U-Net architecture [30] with six layers of depth was utilized for input images downsampled to 256x256 pixels. The model had 3,718,433 parameters and was trained from scratch, with randomly initialized weights, for 120 epochs with a learning rate of $10^{-4}$ and a batch size of 8. A GeForce RTX 3090 NVIDIA graphics processing unit with 24 GB of VRAM was utilized. Using Keras [31] and TensorFlow [32] (versions 2.11.0), the U-Net was trained with the mean squared error (MSE) loss and adaptive moment estimation optimizer. The final model was selected from the epoch with the best validation loss. The detailed model architecture is depicted in Fig. 2B.

Synthetic data from the public datasets were shuffled using Python's (version 3.8.10) built-in function and divided on a train-validation-test split with a 60:20:20 ratio. The data in the train and validation splits were balanced between the two public datasets as follows: For every CT scan, all 10 simulated projections were utilized for Luna16 data, but only the central projection at 0° from PE data was considered. In total, there were 7,302 images for training ($n_{\text{Luna16}} = 4{,}120, n_{\text{PE}} = 3{,}182$ ) and 2,191 images for validating ($n_{\text{Luna16}} = 1{,}130, n_{\text{PE}} = 1{,}061$). Additional data augmentation was implemented with Keras' data generator and applied to the simulated radiographs by randomly shifting the images vertically and horizontally within a range of 0.2 relative to the image dimension and rotating the images within a range of 10°. The model was tested on 1,191 synthetic radiographs generated from public data ($n_{\text{Luna16}} = 131, n_{\text{PE}} = 1{,}060$), and 72 ($n_{\text{Healthy}} = 33, n_{\text{COPD}} = 39$) image pairs corresponding to real and synthetic radiograph pairs from the KRI data. All synthetic test data was retrieved from the central projections at 0°.

## Registration Method and Saliency Maps

For a qualitative comparison of predicted thickness maps, four-level multi-resolution image registration was implemented with the deepali (version 0.6.2) library to register the synthetic to the real thickness map predictions [33]. A stationary velocity field-based transformation model with a stride of nine, a normalized cross-correlation loss, and an adaptive moment estimation optimizer with a learning rate of 0.0001 were utilized.

Saliency maps were generated by adapting the technique described in [34] to the U-Net: The mean over the model gradient for all pixels of each prediction map with respect to the input image was calculated, as suggested by Chatterjee et al. [35].

## Evaluation Metrics and Statistical Analysis

Statistical normality for data distribution was confirmed with the Shapiro-Wilk, D'Agostino's K-squared, and Anderson-Darling tests before calculating the 95% confidence intervals (CI) for mean TLV. Model performance was assessed with the MSE, mean absolute error (MAE), and mean absolute percentage error (MAPE) between the reference standard and predicted TLV for the synthetic and real test data. Pearson correlation coefficient (r) calculation and two-sided Student's t-distribution significance testing were performed. The null hypothesis (r=0) was rejected at a significance level of 0.001 ($P < 0.001$). The statistical analysis was done with SciPy (version 1.10.1) [20]. Bland-Altman analysis and linear fits were calculated with NumPy (version 1.24.3) [36] [37].

# Results

## Dataset Characteristics

The study evaluated the U-Net model on 1,191 synthetic radiographs from public datasets and 72 synthetic and real radiograph pairs from the KRI dataset. The cohort included 33 healthy participants (20 males, mean age [range] = 62 [34, 80]) and 39 participants suffering from COPD (25 males, mean age [range] = 69 [47, 91]). Participants' demographics for the KRI dataset are provided in Table 1.

## Synthetic Radiographs

Quantitative results demonstrating low error rates for TLV and pixel-level thickness estimation for the synthetic radiographs are presented in Table 2. The model is able to predict the TLV for the public and KRI synthetic data with low errors ($MSE_{Public}$ = 0.16 $L^2$, $MAE_{Public}$ = 0.24 L, $MAPE_{Public}$ = 7698/119100 (6.46%); $MSE_{KRI}$ = 0.20 $L^2$, $MAE_{KRI}$ = 0.32 L, $MAPE_{KRI}$ = 420/7200 (5.83%)). A strong correlation exists between the predicted and the reference standard mean TLV and pixel-level thickness values for all test data. For the TLV, Luna16 test data resulted in the highest correlation (r = 0.99, $P < 0.001$) and the lowest error (MSE = 0.09 $L^2$, MAE = 0.23 L, MAPE = 576/13100 (4.40%)). The COPD data resulted in the lowest correlation (r = 0.97, $P < 0.001$) and the highest error (MSE = 0.28 $L^2$, MAE = 0.38 L, MAPE = 254/3900 (6.51%)) for the TLV. The model tends to moderately underestimate the TLV and the pixel-level thickness values.

Fig. 3 shows radiographs with their reference standard and predicted thickness maps for eight example participants from the synthetic public data. There are no considerable qualitative differences between the first four participants corresponding to the Luna16 data and the last four participants corresponding to the PE data. Examples of eight synthetic radiographs from the healthy and COPD KRI data, the corresponding predicted and reference standard lung thickness maps, and their absolute difference are provided in Fig. 4A-D. For both public and KRI data, the predicted thicknesses are somewhat smoother and less detailed than the reference standard thickness maps. Qualitative thickness map differences are more prominent for the KRI data.

## Real Radiographs

TLV estimations for real radiographs are presented in Table 3. Pearson correlation coefficients are slightly lower, and MSE, MAE, and MAPE values are higher for real radiographs in comparison to their synthetic counterparts. Nonetheless, predicted and reference standard TLV values for real radiographs are strongly correlated (r = 0.91, $P < 0.001$). The predicted TLVs for real radiographs are marginally underestimated in comparison to the CT-derived reference standards supplied by the manufacturer.

Table 4 compares the results from our method with the conventional method introduced by Pierce et al. [6] for TLV estimation of real radiographs. TLV values calculated with the Pierce et al. method are overestimated with respect to the CT-derived reference standard values, in contrast to the TLV values estimated with our method. Additionally, wider 95% confidence intervals are reported with the Pierce et al. method in comparison to the proposed method. Given that Pierce et al. utilize the global lung shape information in CXRs for TLV calculation, it is not possible to generate thickness maps to gain lobular or more localized lung volume estimations with this method.

Real radiographs and corresponding predicted thickness maps for four healthy participants and four participants with COPD are shown in Fig. 4E-F. The thickness map predictions from synthetic radiographs in Fig. 4B were registered to the thickness predictions from their real radiograph pairs in Fig. 4F. The registered thickness maps are shown in Fig. 4G. A qualitative comparison of predicted lung thickness maps for the synthetic and real radiograph pairs is provided in Fig. 4H by taking the absolute difference between the real and the registered thickness map predictions. In some cases, the thicknesses were underestimated for the real radiograph of the same participant in Fig. 4F when compared to the reference standard maps generated from CT lung masks in Fig. 4C. Although registration allows for a closer comparison, participant posture differences such as the position of the arms, variations in inspiration cycle, as well as the time-gap between CT and radiograph acquisitions must be considered when comparing real and synthetic radiograph pairs and their predicted thicknesses.

Scatter plots with linear fits of reference standard versus predicted TLV for individual participants are shown for the public test data in Fig. 5A, the synthetic KRI data in Fig. 5B, and the real KRI data in Fig. 5C. For the KRI data, synthetic radiographs had a better linear fit (slope = 1.05 [0.99, 1.11], offset = -0.21 [-0.56, 0.13] L), and a larger correlation coefficient (r = 0.97, $P < 0.001$) than their real radiograph pairs (slope = 1.14 [1.02, 1.26], offset = -0.57 [-1.21, 0.08] L; r = 0.91, $P < 0.001$). The best linear fit (slope = 1.01 [1.00, 1.02], offset = -0.02 [-0.07, 0.03] L) and the strongest correlation (r = 0.99, $P < 0.001$) were given by synthetic radiographs from the public datasets. The Bland-Altman plots for the corresponding scatter plots in Fig. 5A-C are depicted in Fig. 5D-F. The estimated volumes for 95% of participants were within ± 0.77 L for the synthetic public radiographs, ± 0.87 L for the synthetic KRI radiographs, and ± 1.13 L for the real KRI radiographs. Overall, there is no observable systematic bias as the mean of differences for synthetic ($mean_{Public}$ = -0.04 L, $mean_{KRI}$ = -0.07 L) and real (mean = -0.13 L) radiographs are close to zero. The linear regression line of differences suggests that the proportional error observed is minimal for the synthetic radiographs ($slope_{Public}$ = -0.04 [-0.05, -0.03], $offset_{Public}$ = 0.13 [0.09, 0.17] L; $slope_{KRI}$ = -0.10 [-0.15, -0.05], $offset_{KRI}$ = 0.48 [0.18, 0.77] L) and minor for the real radiographs (slope = -0.28 [-0.36, -0.20], offset = 1.30 [0.87, 1.73] L).

The model performance was separately assessed for four real radiographs that included foreign objects. These cases are depicted in Fig. 6, showing the real radiographs and the predicted thickness maps, two of which correspond to healthy participants and two to participants with COPD. In all cases, the strongly absorbing foreign object present in the scan causes changes to the intensity range in the radiograph, which in turn results in failed thickness map predictions by the model. These participants were excluded from the KRI data, as shown in Fig. 1B, and were only investigated for the failure analysis in Fig. 6.

The overlay of the resulting saliency maps over the corresponding synthetic and real radiographs from the KRI dataset are depicted in Fig. S1: The mean over 33 healthy and 39 COPD participants are shown in Fig. S1A-B, and an example radiograph from each group is presented in Fig. S1C-D, respectively. Fig. S1E demonstrates failed thickness estimation cases in the presence of foreign bodies for real radiographs. In Fig. S1A-D, it can be observed that the model utilizes landmarks such as the spine, the heart, the curvatures of the ribs, and the outline of the lungs to navigate the lobes and focus on the area inside the lungs. The presence of foreign bodies alters the saliency maps, causing a loss of focus on some key landmarks, as seen in Fig. S1E. This indicates that the model struggles to recognize these landmarks when foreign bodies are present, consequently affecting its performance.

## Discussion

This study investigated pixel-level lung thickness map and TLV estimation using a U-Net trained on synthetic radiographs and applied to real radiographs. Key findings demonstrated low error rates in TLV predictions: synthetic public data MSE=0.16 $L^2$, synthetic KRI data MSE=0.20 $L^2$, and real radiographs MSE=0.35 $L^2$. Strong correlations were measured between the predicted and reference standard TLV values for synthetic ($r_{Public} = 0.99$, $P < 0.001$; $r_{KRI} = 0.97$, $P < 0.001$) and real radiographs ($r = 0.91$, $P < 0.001$).

Larger positive Pearson's correlation coefficients and lower MSE, MAE, and bias showed that the model was able to better predict TLV values for the synthetic public test data in comparison to the synthetic KRI test data. This behavior is explainable as the model train and validation data splits were also drawn from the same public data distribution. Therefore, it is understandable that the model would have sub-par performance on the external synthetic data taken from the KRI data distribution.

We examined the robustness of our method for multiple lung pathologies, namely lung nodules, PE, and COPD, in comparison to healthy lungs. For both synthetic and real data, regardless of the pathology, our reference standard and predicted TLV results were within previously reported CT-derived mean TLV range for healthy lungs as described by Haas et al. [38] (n=302, mean ± standard deviation = 5.28±0.95 L) and Wisselink et al. [39] ($n_{Female}$=151, mean ± standard deviation = 4.7±0.9 L; $n_{Male}$=139, mean ± standard deviation = 6.2±1.2 L). Our findings for the PE test set fell towards the lower end of this range, which can be attributed to the inclusion of acute and chronic PE types in the RSNA PE dataset [18], given participants with acute PE experience a reduction of pulmonary function and capacity [40].

Compared to previous work on lung volume estimation with deep-learning models for real radiographs by Sogancioglu et al. [7] (MAE = 0.59 L, r = 0.86), our model achieved better results (MAE = 0.46 L, r = 0.91). Additionally, we are able to obtain not only the total lung volume but also a pixel-level lung thickness map. Since such a lung thickness map cannot be calculated from real radiographs due to the missing reference standard information, the training must be performed exclusively with simulated radiographs calculated from CT scans. Here, it should be emphasized that we are still able to test the model on real, non-simulated, radiographs and report promising results based on an evaluation with respect to the CT-derived total lung volume.

However, transferring knowledge from CT scans to real radiographs presents several inherent hurdles, mainly different participant postures and varying inspiration levels in CT data [7]. Advances in the field of image registration could enable more accurate translation from synthetic to real radiographs, thereby reducing the effects of participant posture and inspiration on thickness map prediction and TLV calculation. Additionally, neglected physical effects such as Compton scattering could also account for differences between real and synthetic radiographs. Yet, our results for synthetic radiographs indicate that with further consideration of these effects, even lower predicted TLV errors can be achieved for real radiographs.

The proposed method could supplement or substitute conventional approaches for lung volume estimation. When CT imaging is ruled out due to cost, availability, or excessive radiation exposure concerns, CXRs are a common alternative, for example, in repeated follow-up screenings. An automated per-pixel analysis of the TLV will allow for a better understanding of pathologies that alter the lung volume heterogeneously, such as in the cases of emphysema and pulmonary fibrosis [41]. Specifically, such thickness maps will serve as a quantitative measure of localized and lobular volume changes, assisting physicians in diagnosing and preventing the progression of pulmonary diseases during routine CXR screenings. Furthermore, the proposed method has potential applications in pulmonary resection surgery and emerging novel imaging techniques [12, 14-16].

The study had several limitations. First, the sample of real radiographs was relatively small, restricted to healthy and participants with COPD from a single medical center. Second, the unprocessed radiographs were selected from the KRI setup to best match synthetically generated radiograph pairs to avoid the need for vendor-specific post-processing algorithms, which are closed sources and unavailable from the imaging device manufacturers. This restricted the results of our retrospective study to healthy and COPD participants scanned at the KRI setup. Future works should validate the applicability of these findings to other restrictive lung diseases. Furthermore, the current workflow could benefit from additional improvements. For example, deep learning-based segmentation can be used for different tissue types instead of HU thresholding. Moreover, complementing the proposed method with other techniques based on pulmonary function data, such as the one presented by Boulogne et al. [12], would allow for a comprehensive estimation of various lung volumes.

Our study demonstrates that pixel-level lung thickness maps for real and synthetic radiographs can be generated with a U-Net trained on synthetic CT-derived radiographs for subsequent calculation of the TLV. Low mean squared and mean absolute errors were achieved, and strong correlations were observed between the CT-derived reference standard and the U-Net predicted TLV values obtained from the generated thickness maps. Future research could investigate the additional use of lateral radiographs for thickness map estimation and explore other network architectures.

**Tables**

**Table 1**

**Participant Demographics for Klinikum Rechts der Isar Dataset (n=72)**

| Parameter \ Subset | Healthy | COPD |
|---|---|---|
| Male | 20 (28%) | 25 (35%) |
| Female | 13 (18%) | 14 (19%) |
| Age (years) | 62 [34, 80] | 69 [47, 91] |

Note. Data are number of participants with percentage in parentheses or mean with range in brackets.

COPD = Chronic Obstructive Pulmonary Disease

**Table 2**

**Total Lung Volume and Pixel-level Thickness for Synthetic Radiographs**

| Dataset \ Volume | Luna16 (n = 131) | PE (n = 1,060) | Healthy (n = 33) | COPD (n = 39) |
|---|---|---|---|---|
| Mean Reference Standard [95% CI] (L) | 5.45 [5.00 – 5.90] | 4.10 [3.96 – 4.24] | 5.12 [4.51 – 5.73] | 5.79 [5.13 – 6.44] |
| Mean Predicted [95% CI] (L) | 5.39 [4.60 – 5.83] | 4.06 [3.93 – 4.20] | 5.13 [4.55 – 5.72] | 5.64 [5.04 – 6.24] |
| MSE ($L^2$) | 0.09 | 0.17 | 0.11 | 0.28 |
| MAE (L) | 0.23 | 0.24 | 0.25 | 0.38 |
| MAPE | 576/13100 (4.40%) | 7122/106000 (6.72%) | 166/3300 (5.03%) | 254/3900 (6.51%) |
| Pearson Coefficient | 0.99 (*P* < 0.001) | 0.99 (*P* < 0.001) | 0.98 (*P* < 0.001) | 0.97 (*P* < 0.001) |
| Dataset \ Thickness | Luna16 (n = 131) | PE (n = 1,060) | Healthy (n = 33) | COPD (n = 39) |
| Mean Reference Standard [95% CI] (mm) | 85.40 [83.41 – 87.40] | 71.63 [70.88 – 72.39] | 79.36 [76.15 – 82.56] | 87.33 [82.90 – 91.76] |
| Mean Predicted [95% CI] (mm) | 84.55 [82.66 – 86.44] | 70.83 [70.12 – 71.55] | 79.43 [76.71 – 82.16] | 85.05 [81.38 – 88.73] |
| MSE ($mm^2$) | 20.17 | 36.56 | 28.50 | 55.36 |
| MAE (mm) | 3.68 | 4.24 | 4.11 | 5.75 |
| MAPE | 580/13100 (4.43%) | 8767/106000 (8.27%) | 171/3300 (5.18%) | 262/3900 (6.72%) |
| Pearson Coefficient | 0.92 | 0.88 | 0.80 | 0.85 |
| | ($P < 0.001$) | ($P < 0.001$) | ($P < 0.001$) | ($P < 0.001$) |

COPD = Chronic Obstructive Pulmonary Disease

MAE = Mean Absolute Error

MAPE = Mean Absolute Percentage Error

MSE = Mean Squared Error

PE = Pulmonary Embolism

**Table 3**

**Total Lung Volume for Real Radiographs**

| Dataset \ Metric | Healthy (n = 33) | COPD (n = 39) | All (n = 72) |
|---|---|---|---|
| Mean Reference Standard [95% CI] (L) | 4.85 [4.43 – 5.26] | 5.51 [5.05 – 5.97] | 5.21 [4.89 – 5.52] |
| Mean Predicted [95% CI] (L) | 4.83 [4.47 – 5.19] | 5.28 [4.93 – 5.64] | 5.08 [4.82 – 5.33] |
| MSE ($L^2$) | 0.30 | 0.39 | 0.35 |
| MAE (L) | 0.43 | 0.49 | 0.46 |
| MAPE | 304/3300 (9.21%) | 354/3900 (9.08%) | 658/7200 (9.14%) |
| Pearson Coefficient | 0.88 ($P < 0.001$) | 0.92 ($P < 0.001$) | 0.91 ($P < 0.001$) |

COPD = Chronic Obstructive Pulmonary Disease

MAE = Mean Absolute Error

MAPE = Mean Absolute Percentage Error

MSE = Mean Squared Error

**Table 4**

**Comparison of Total Lung Volume Estimation Methods for Real Radiographs**

| Dataset \ Method | Healthy (n = 33) | COPD (n = 39) | All (n = 72) |
|---|---|---|---|
| Pierce et al. | 6.57 | 7.61 | 7.13 |
| Mean [95% CI] (L) | [5.91 – 7.23] | [7.05 – 8.17] | [6.70 – 7.57] |
| U-Net (ours) | 4.83 | 5.28 | 5.08 |
| Mean [95% CI] (L) | [4.47 – 5.19] | [4.93 – 5.64] | [4.82 – 5.33] |
| CT (manufacturer) | 4.85 | 5.51 | 5.21 |
| Mean [95% CI] (L) | [4.43 – 5.26] | [5.05 – 5.97] | [4.89 – 5.52] |

COPD = Chronic Obstructive Pulmonary Disease

**Figures**

**Figure 1.** Data selection flowchart. **A)** Public chest CT data from the Luna16 (2016) and the RSNA pulmonary embolism (PE) challenges (2020) were considered to generate synthetic radiographs. Only scans acquired with 120kVp with a determined diagnosis were selected. Luna16 and PE data were split separately on a 60:20:20 ratio for the train, validation, and test sets. Synthetic radiographs were generated by simulating 10 projections for each sample from -10° to 10° at 2° steps. The following was implemented to balance the two datasets for the training process: All projections were utilized for the train and validation sets from the Luna16 data. Only central projections at 0° were chosen for the Luna16 test set, and all of the PE data sets, as this angle provides the most accurate real radiograph representation. **B)** Klinikum Rechts der Isar (KRI) chest CT scans [16] were retrospectively selected (October 2018-December 2019), such that all scans were acquired at 120kVp and had a corresponding radiograph scanned no longer than a week apart. Synthetic radiographs were generated by simulating one projection for each CT scan at 0°. The synthetic and the real radiographs were kept as an external test set.

**Figure 2.** Illustrated workflow for generating synthetic radiographs and lung thickness maps for training a U-Net. **A)** The CT scan is segmented into soft tissue, adipose tissue, and bone. An X-ray spectrum is simulated for the forward projection of the CT to generate the synthetic radiograph. To create the two-dimensional reference standard lung thickness density map, the three-dimensional lung segmentation mask is forward projected. The U-Net model is trained with synthetic radiograph input data and corresponding thickness map reference standard labels. **B)** The U-Net model architecture for 256-by-256 pixel input data. The number of feature channels is given above the corresponding block. A three-by-three convolution links a feature block to the next, if not specified otherwise. Upsampling was implemented with nearest neighbor interpolative resizing.

**Figure 3.** Qualitative results of eight frontal chest radiographs from the synthetic public test set, where each column represents one participant. Participants one through four are from the Luna16 test set, and participants five through eight are from the pulmonary embolism test set. **A)** Simulated input radiographs, **B)** predicted thickness maps, **C)** reference standard thickness maps, and **D)** absolute difference between the reference standard and predicted maps. The color bars for B) and C) indicate lung thickness in millimeters [mm]. The color bar for D) indicates the pixel-level estimation error in mm.

**Figure 4.** Qualitative results of eight frontal chest radiographs from the Klinikum Rechts der Isar test set, where each column represents one participant. The ages of participants are provided in parenthesis. Participant 1 is female, and all other participants are male. Participants one through four are healthy, and participants five through eight suffer from chronic obstructive pulmonary disease. **A)** Simulated input radiographs, **B)** predicted thickness maps for A), **C)** reference standard thickness maps for A), **D)** absolute difference between reference standard C) and predicted B) thickness maps, **E)** real input radiographs, and **F)** predicted thickness maps for E), **G)** registration of the synthetic B) to the real F) thickness map predictions, **H)** absolute differences in the registered G) and the real F) thickness map predictions. The CT scan and the lung segmentation were forward projected respectively to obtain A) and C). The predicted thickness map for real radiographs F) and the registered predicted thickness map for their synthetic pairs G) can be compared for each case in H). The color bars for B), C), F), and G) indicate lung thickness in millimeters [mm]. The color bars for D) and H) indicate the pixel-level estimation error in mm.

**Figure 5**. Scatter plots for the reference standard versus predicted total lung volumes in liters [L] and Bland-Altman plots for the difference of predicted and reference standard over the reference standard total lung volumes in liters [L], respectively, for **A)** and **D)** synthetic radiographs from the public test sets of the Luna16 (orange) and the RSNA pulmonary embolism (PE)(blue) challenges, **B)**

and **E)** synthetic radiographs from the Klinikum Rechts der Isar dataset, including healthy participants (blue), and participants suffering from chronic obstructive pulmonary disease (COPD)(orange), and **C)** and **F)** real radiograph pairs for the same participants in B) and E). The total number of radiographs is given by *n*. Pearson's correlation coefficient, *r*, and statistical significance are reported. The linear fit is depicted with the black line, and the 95% confidence intervals (CI) are reported in the legends.

**Figure 6.** Qualitative results of four real frontal chest radiographs with failed thickness map predictions, where each column represents one participant. **A)** input radiographs and **B)** corresponding predicted thickness maps. For all participants, the changes in the intensity range caused by strongly absorbing features in the input radiograph resulted in failed predictions. The color bar for B) indicates lung thickness in millimeters [mm]. Radiographs one and two correspond to healthy participants. Radiographs three and four correspond to participants suffering from chronic obstructive pulmonary disease.

**Figure S1.** Overlay of saliency maps with the corresponding synthetic and real radiographs from the Klinikum Rechts der Isar dataset. **A-B)** The mean over 33 healthy participants and 39 participants suffering from chronic obstructive pulmonary disease (COPD) for synthetic and real radiographs, respectively, **C-D)** representative examples from A) and B) respectively, and **E)** representative examples of failed thickness estimation cases for real radiographs of healthy and COPD participant in the presence of foreign bodies.

**Figure 1.**

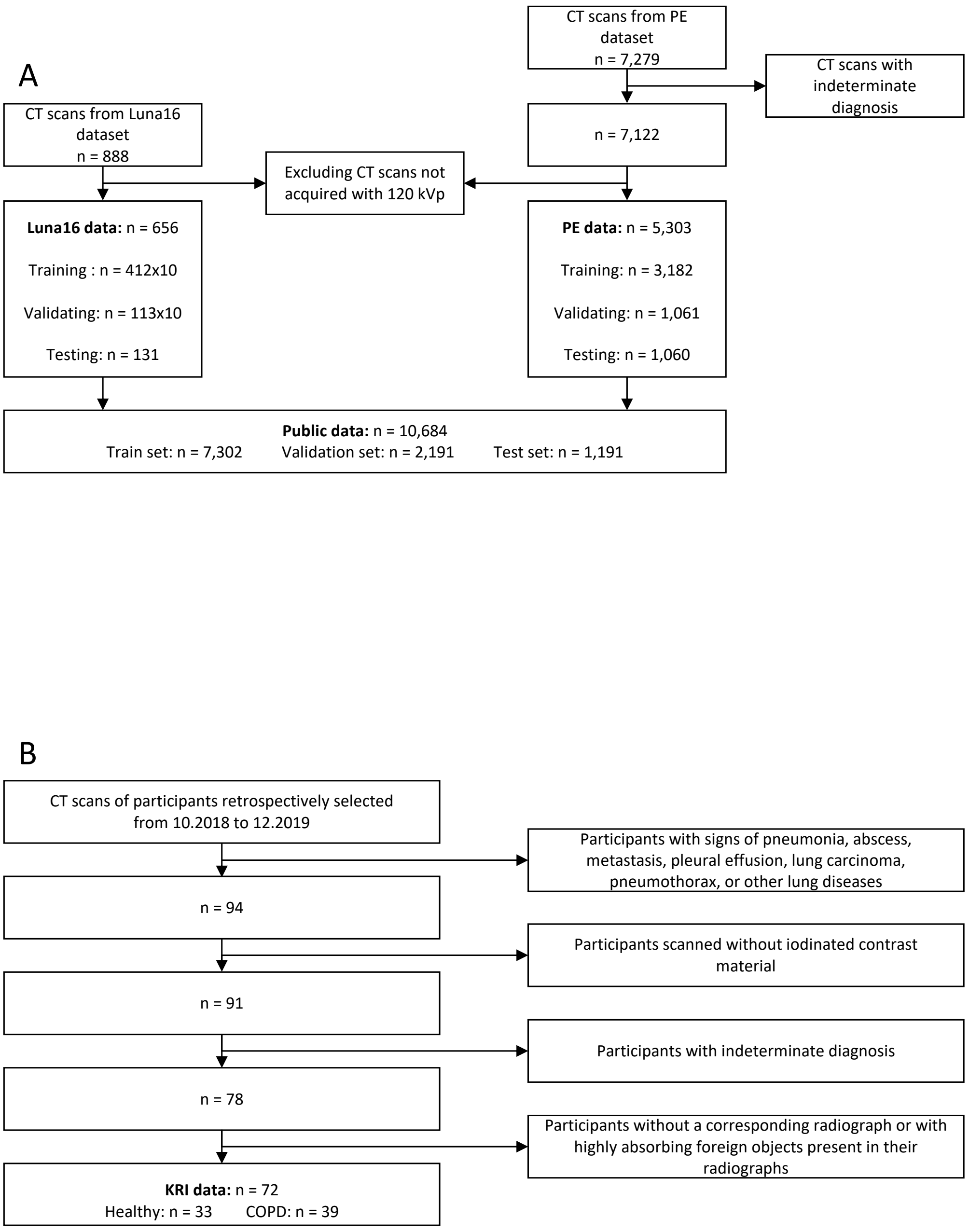
A
CT scans from PE dataset n = 7,279
CT scans with indeterminate diagnosis
CT scans from Luna16 dataset n = 888
n = 7,122
Excluding CT scans not acquired with 120 kVp
Luna16 data: n = 656
Training : n = 412x10
Validating: n = 113x10
Testing: n = 131
PE data: n = 5,303
Training: n = 3,182
Validating: n = 1,061
Testing: n = 1,060
Public data: n = 10,684
Train set: n = 7,302
Validation set: n = 2,191
Test set: n = 1,191
B
CT scans of participants retrospectively selected from 10.2018 to 12.2019
Participants with signs of pneumonia, abscess, metastasis, pleural effusion, lung carcinoma, pneumothorax, or other lung diseases
n = 94
Participants scanned without iodinated contrast material
n = 91
Participants with indeterminate diagnosis
n = 78
Participants without a corresponding radiograph or with highly absorbing foreign objects present in their radiographs
KRI data: n = 72
Healthy: n = 33
COPD: n = 39

**Figure 2.**

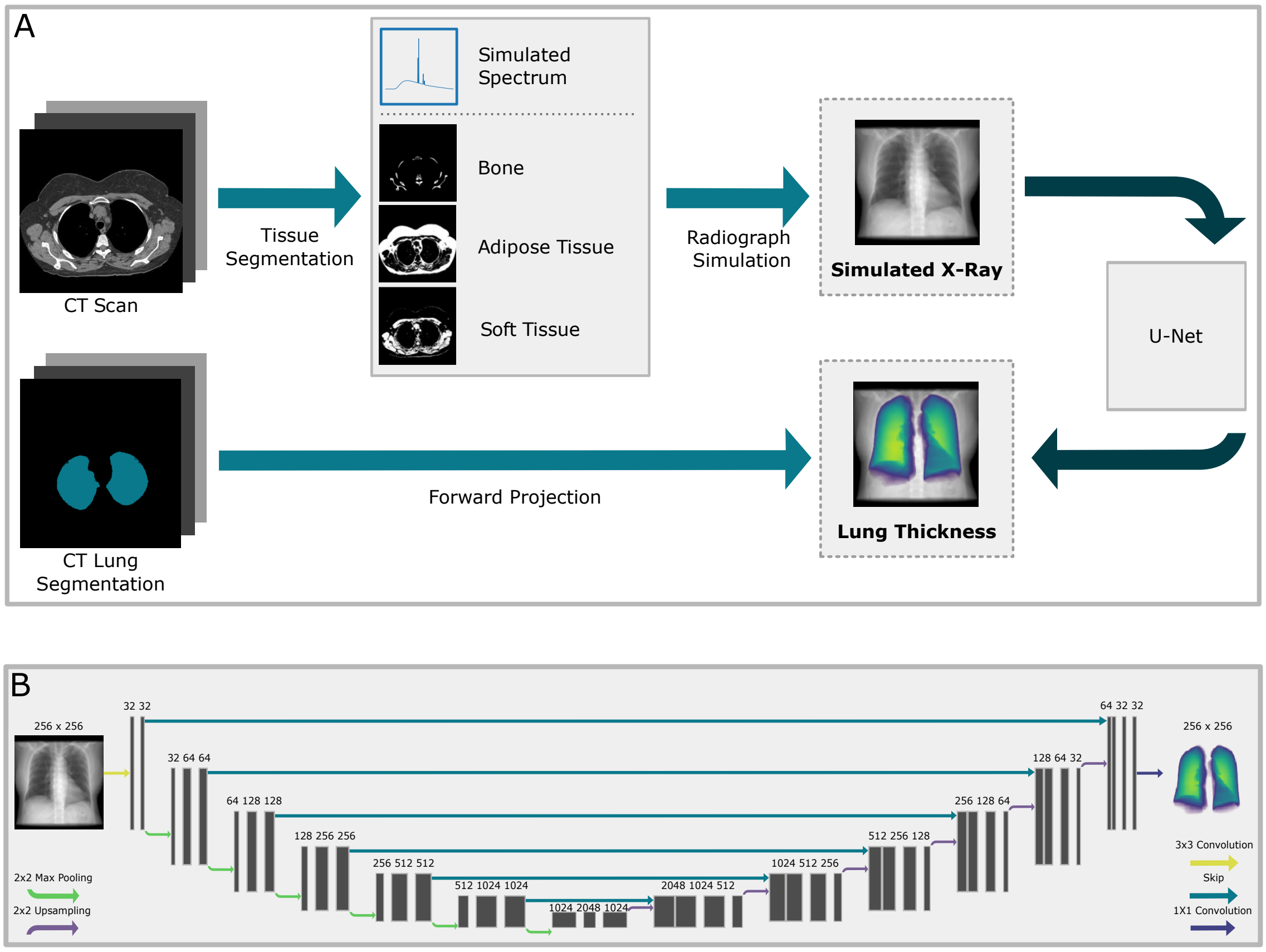
A
Simulated Spectrum
Bone
Adipose Tissue
Soft Tissue
CT Scan
Tissue Segmentation
Radiograph Simulation
Simulated X-Ray
U-Net
Lung Thickness
Forward Projection
CT Lung Segmentation
B
256 x 256
32 32
32 64 64
64 128 128
128 256 256
256 512 512
512 1024 1024
1024 2048 1024
2048 1024 512
1024 512 256
512 256 128
256 128 64
128 64 32
64 32 32
256 x 256
2x2 Max Pooling
2x2 Upsampling
3x3 Convolution
Skip
1X1 Convolution

**Figure 3.**

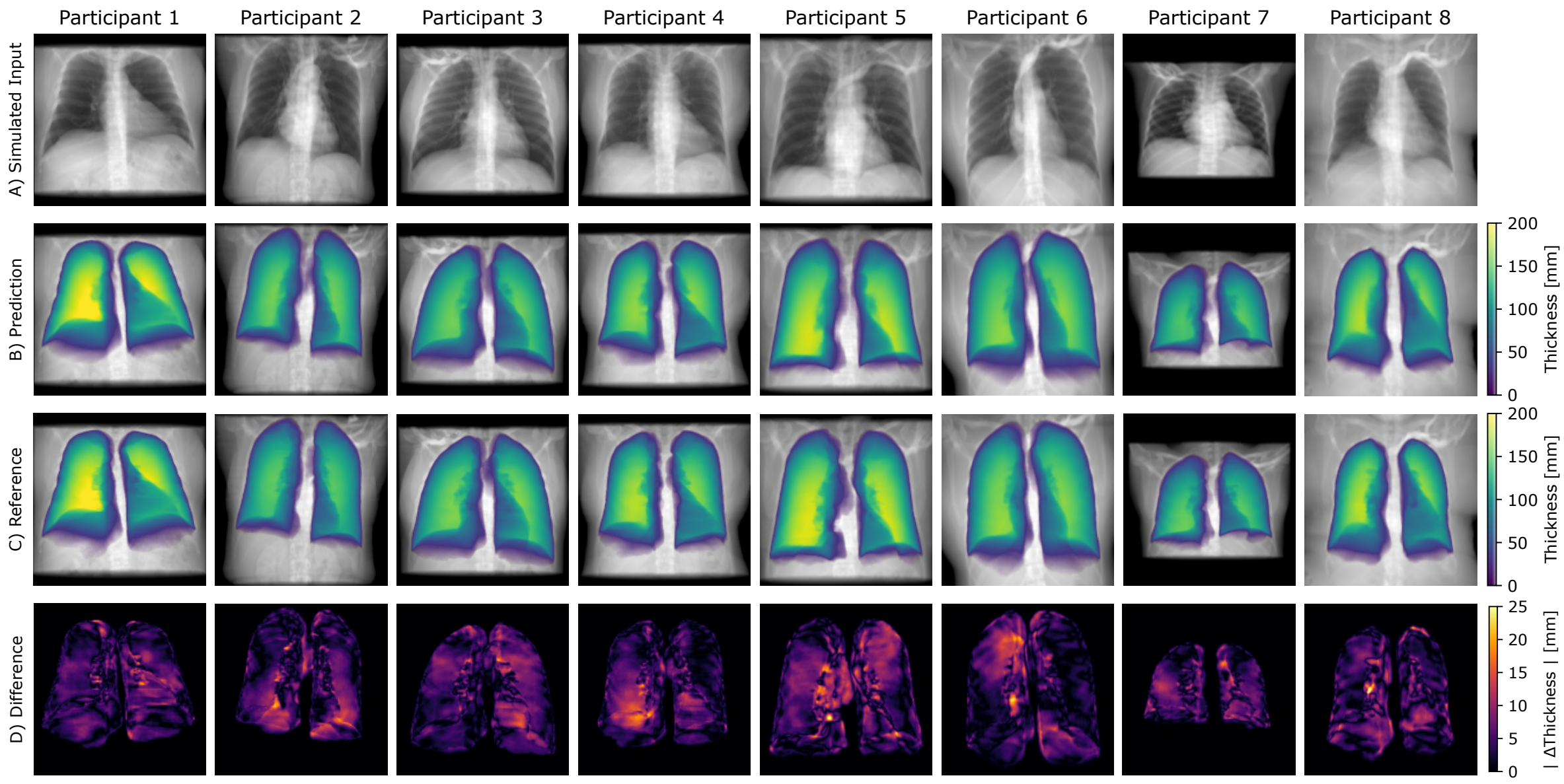

Participant 1
Participant 2
Participant 3
Participant 4
Participant 5
Participant 6
Participant 7
Participant 8
A) Simulated Input
B) Prediction
C) Reference
D) Difference
200
150
100
50
0
Thickness [mm]
200
150
100
50
0
Thickness [mm]
25
20
15
10
5
0
| ΔThickness | [mm]

**Figure 4.**

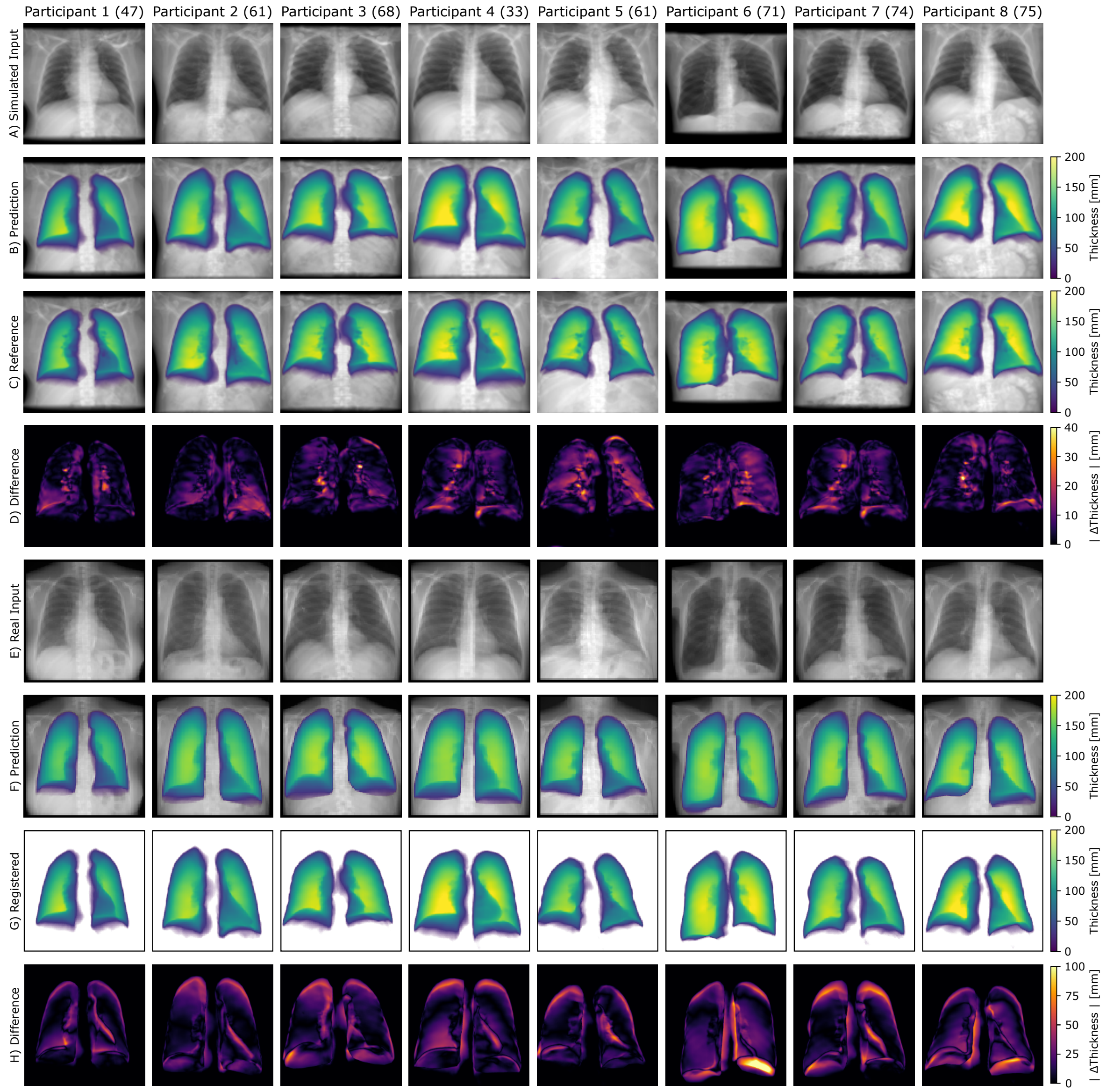

Participant 1 (47)
Participant 2 (61)
Participant 3 (68)
Participant 4 (33)
Participant 5 (61)
Participant 6 (71)
Participant 7 (74)
Participant 8 (75)
A) Simulated Input
B) Prediction
C) Reference
D) Difference
E) Real Input
F) Prediction
G) Registered
H) Difference
Thickness [mm]
| ΔThickness | [mm]
0
50
100
150
200
0
10
20
30
40
0
25
50
75
100

**Figure 5.**

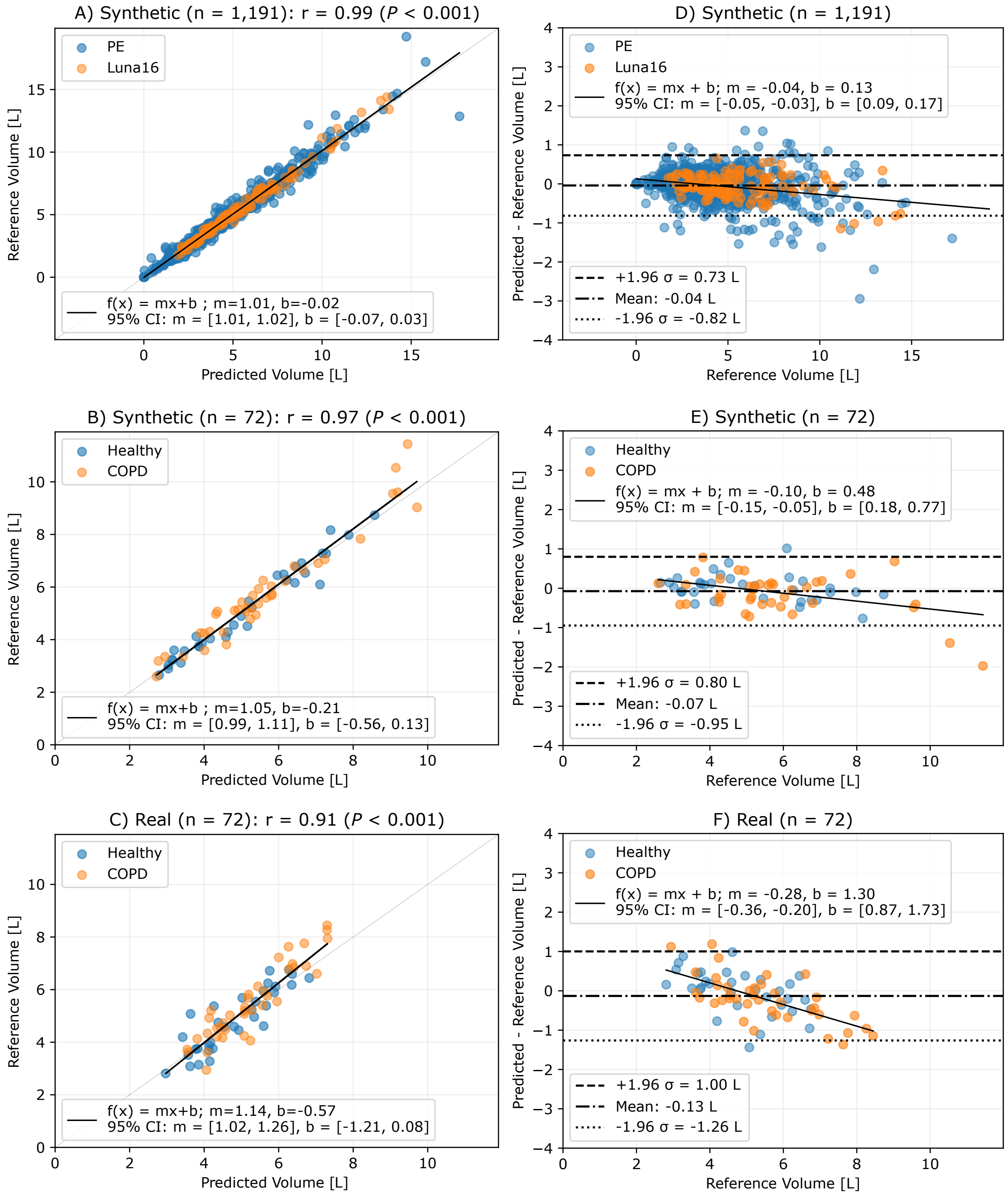
A) Synthetic (n = 1,191): r = 0.99 (P < 0.001)
PE
Luna16
f(x) = mx+b ; m=1.01, b=-0.02
95% CI: m = [1.01, 1.02], b = [-0.07, 0.03]
Reference Volume [L]
Predicted Volume [L]
D) Synthetic (n = 1,191)
PE
Luna16
f(x) = mx + b; m = -0.04, b = 0.13
95% CI: m = [-0.05, -0.03], b = [0.09, 0.17]
+1.96 σ = 0.73 L
Mean: -0.04 L
-1.96 σ = -0.82 L
Predicted - Reference Volume [L]
Reference Volume [L]
B) Synthetic (n = 72): r = 0.97 (P < 0.001)
Healthy
COPD
f(x) = mx+b ; m=1.05, b=-0.21
95% CI: m = [0.99, 1.11], b = [-0.56, 0.13]
Reference Volume [L]
Predicted Volume [L]
E) Synthetic (n = 72)
Healthy
COPD
f(x) = mx + b; m = -0.10, b = 0.48
95% CI: m = [-0.15, -0.05], b = [0.18, 0.77]
+1.96 σ = 0.80 L
Mean: -0.07 L
-1.96 σ = -0.95 L
Predicted - Reference Volume [L]
Reference Volume [L]
C) Real (n = 72): r = 0.91 (P < 0.001)
Healthy
COPD
f(x) = mx+b; m=1.14, b=-0.57
95% CI: m = [1.02, 1.26], b = [-1.21, 0.08]
Reference Volume [L]
Predicted Volume [L]
F) Real (n = 72)
Healthy
COPD
f(x) = mx + b; m = -0.28, b = 1.30
95% CI: m = [-0.36, -0.20], b = [0.87, 1.73]
+1.96 σ = 1.00 L
Mean: -0.13 L
-1.96 σ = -1.26 L
Predicted - Reference Volume [L]
Reference Volume [L]

**Figure 6.**

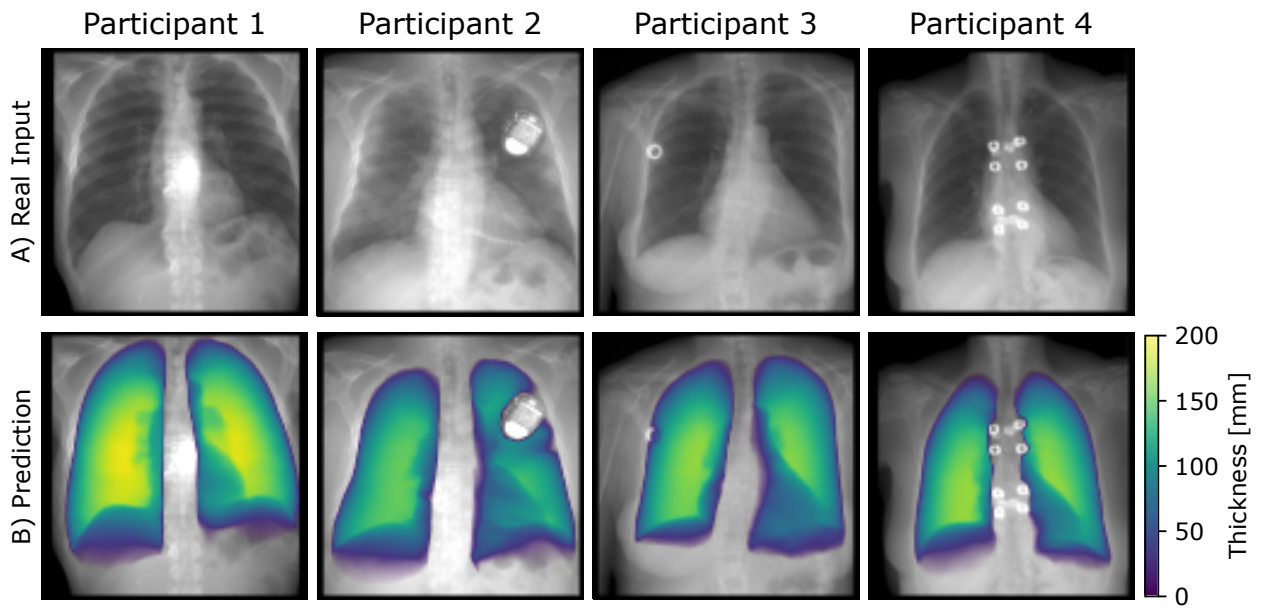

Participant 1
Participant 2
Participant 3
Participant 4
A) Real Input
B) Prediction
200
150
100
50
0
Thickness [mm]

**Figure S1.**

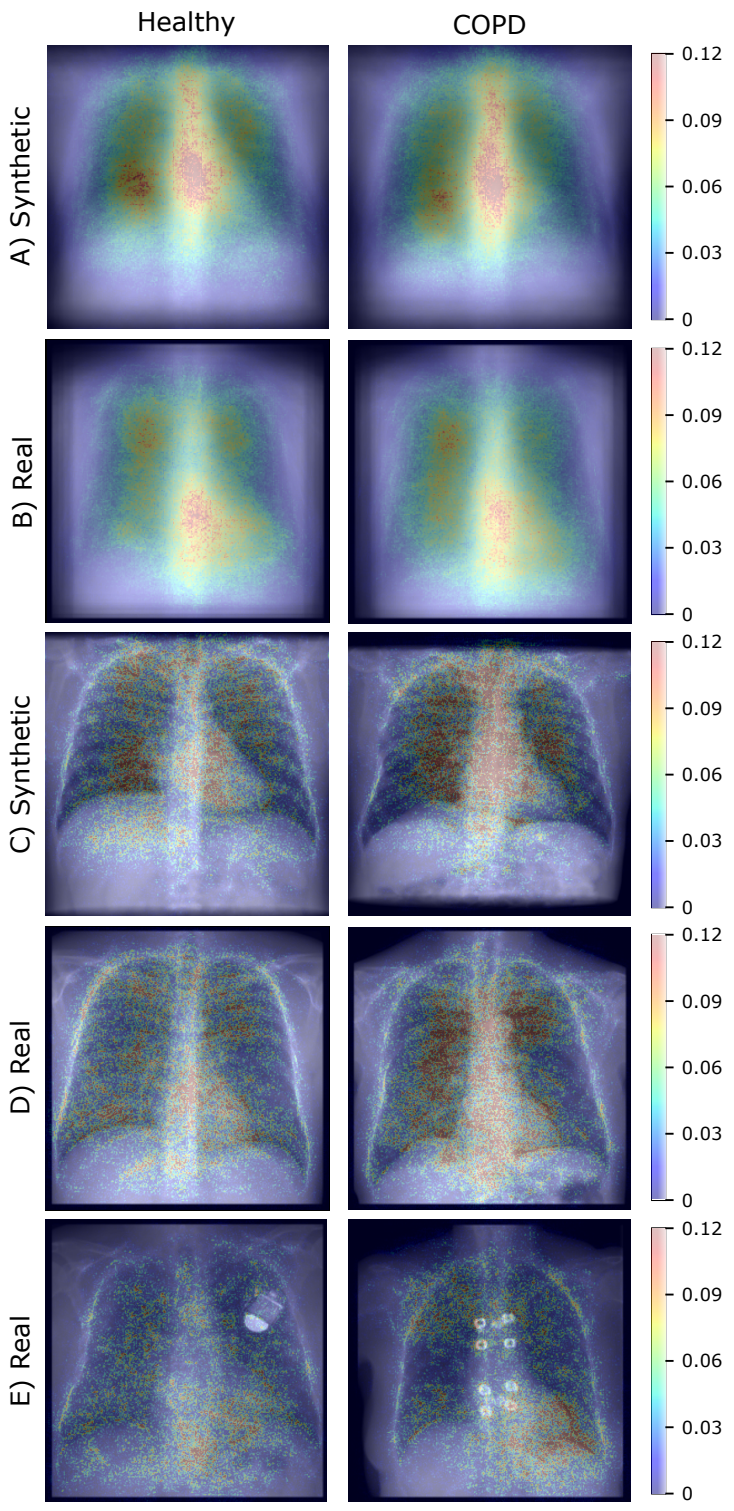

Healthy
COPD
A) Synthetic
B) Real
C) Synthetic
D) Real
E) Real
0.12
0.09
0.06
0.03
0

**Table S1**

**CT Reconstruction Kernel and Contrast Agent Information**

| Dataset \ CT Count | Luna16 | PE | KRI |
|---|---|---|---|
| Smooth Kernel | 123 (19%) | 5,303 (100%) | 0 |
| Standard Kernel | 319 (49%) | 0 | 0 |
| Sharp Kernel | 214 (32%) | 0 | 72 (100%) |
| Non-conrast | 520 (79%) | 0 | 0 |
| Contrast-enhanced | 136 (21%) | 5,303 (100%) | 72 (100%) |
| [Phase] | [Unknown] | [Pulmonary Arterial] | [Pulmonary Arterial] |

Data are number of participants with percentage in parentheses.